\documentclass[aps,twocolumn, amsmath,nofootinbib]{revtex4-1}
\usepackage{amssymb}

\def\be{\begin{equation}}
\def\ee{\end{equation}}
\def\bea{\begin{eqnarray}}
\def\eea{\end{eqnarray}}

\newcommand{\eq}[1]{Eq.~(\ref{#1})}

\begin{document}
\author{Mahdiyar~Noorbala}

\email{noorbala@stanford.edu}

\author{Vitaly~Vanchurin}

\email{vanchurin@stanford.edu}

\affiliation{Department of Physics, Stanford University, Stanford, CA 94305}

\title{Geocentric cosmology: a new look at the measure problem}

\begin{abstract}

We show that most of cutoff measures of the multiverse violate some of the basic properties of probability theory when applied repeatedly to predict the results of local experiments. Starting from minimal assumptions, such as Markov property, we derive a correspondence between cosmological measures and quantum field theories in one lesser dimension. The correspondence allows us to replace the picture of an infinite multiverse with a finite causally connected region accessible by a given observer in conjunction with a Euclidean theory defined on its past boundary. 

\end{abstract}

\maketitle

\section{Introduction}

The old problem of initial conditions in the big bang cosmology was replaced by a measure problem in the inflationary cosmology. In order to make predictions in eternally inflating multiverse \cite{Vilenkin83,Linde1982,Guth1980,Linde1986} one usually postulates a regularization scheme, or a measure \cite{Linde94,Bousso,Linde07,Winitzki, SGLNSV,Vanchurin,VVW,GSVW,GV2008}, which could be repeatedly tested against observations. In this paper we show that most of the cutoff measures are inconsistent with the most basic properties of the probability theory when applied to local experiments.  To resolve the issue we propose a new approach to the measure problem which allows us to explore more easily the space of consistent measures. 

The paper is organized as follows. In the next section we discuss the inconsistencies of cosmological measure which violate the basic principles of the probability theory. In Sec. III we provide a detailed analysis of the local experiments which motivates a new approach to the measure problem discussed in Sec. IV. The main results are summarized in the conclusion. 

\section{Cutoff Measures}

In this section we show an inconsistency in the predictions of cutoff measures about local experiments.  Let us consider a semi-classical picture of eternal inflation.  We can think of the multiverse as a realization of a random process of formation of pocket universes with rates determined by quantum processes such as tunneling or fluctuations.  In addition, other random processes take place within each pocket universe which in our discussion amount to local experiments in local labs.

We define a local lab as a spacetime region $R$ which is small enough compared with the curvature scale to be safely approximated by a flat metric.  (Technically our results hold for any lab satisfying this definition; but in practice we only use labs extended across regions much smaller than the horizon as well as the curvature scale.)  The boundary of $R$ can be split into a past $R^-$ and a future $R^+$ such that no causal signal can reach $R$ without passing through $R^-$.  In other words, a knowledge of the initial state on $R^-$ is enough to determine the state everywhere on $R$.\footnote{More precisely, $R^-$ is the smallest subset of the boundary of $R$ such that its future domain of dependence $D^+(R^-)$ contains $R$.}  We concentrate on those (double) experiments in which a particular initial measurement is performed on a system at the surface $R^-$ with outcome $\phi_1$ and another final measurement at $R^+$ with outcome $\phi_2$.  Examples of such regions are: (i) causal diamonds with $R^-$ being the future light-cone and $R^+$ being the past light-cone; (ii) two spacelike surfaces $R^\pm$ joined at their boundaries; and (iii) hypercube with eight surfaces, seven forming $R^-$ and one being $R^+$.  In the last case, $R^-$ is not a conventional equal-time surface of initial conditions but it doesn't forbid us from propagating the initial knowledge forward.  We may regard the information on the six ``walls'' in two ways: either as part of the initial conditions set by measurement, or as fixed (reflecting or absorbing) boundary conditions that shield the experiment from the outside world.

Since the total number of such labs is infinite we use a cutoff measure which regularizes the infinite four-volume of spacetime and results in finite probabilities.  By cutoff measure we mean, except when explicitly stated otherwise, one that is defined via a global time cutoff surface.  It's also assumed, as is typical in the literature,  that the spatial direction is regulated by picking a finite comoving initial spacelike surface that remains of finite size at any given time and infinities arise only at future infinity. It can be shown that the result is almost surely independent on the choice of initial surface \cite{Linde94}. Of course, it would be possible to alter predictions by changing the way the spatial direction is regulated, but we will not do that here.

Suppose that a cutoff measure is given.  We can divide the regularized four-volume of spacetime into box regions of fixed size $\ell$ (which characterizes the region $R$) and count regions with various $\phi_{1,2}$.  We then calculate the following probability densities (or just probabilities for short) by taking the cutoff to infinity:\footnote{We assume that these quantities do not depend on the arbitrariness that exists in tiling the spacetime with boxes, except for ${\cal O}(1)$ factors.  To remove this ${\cal O}(1)$-ambiguity, we can alternatively do the counting by a volume-weighted integral over all possible box centers.}

\begin{itemize}

\item $\mu(\phi_1)$: The probability that a local experiment, confined to a box of size $\ell$ in the multiverse, has outcome $\phi_1$ in the initial measurement.

\item $M(\phi_2, \phi_1)$: The joint probability that a local experiment, confined to a box of size $\ell$ in the multiverse, has outcomes $\phi_1$ and $\phi_2$ in the initial and final measurement, respectively.

\item $M(\phi_2 | \phi_1)$: The conditional probability that a local experiment, confined to a box of size $\ell$ in the multiverse and with initial measurement outcome $\phi_1$, has outcome $\phi_2$ in the final measurement.

\end{itemize}

Evidently, all three functions $\mu$, joint $M(\cdot, \cdot)$ and conditional $M(\cdot | \cdot)$ depend on the cutoff measure from which they are derived and on $\ell$ through the dependence of the domain of $\phi$ on $\ell$.  Naively one may expect
\begin{equation} \label{M=muM}
M(\phi_2, \phi_1) = \mu(\phi_1) M(\phi_2 | \phi_1),
\end{equation}
which implies
\begin{equation} \label{mu=intM}
\mu(\phi_1) = \int d\phi_2 M(\phi_2, \phi_1),
\end{equation}
to hold true; but actually neither of Eqs.~(\ref{M=muM}) and (\ref{mu=intM}) are correct in general.  The reason is that the two $M$ functions (joint and conditional) search for whole regions $R$ whereas $\mu$ searches for past boundaries $R^-$ only.  However, below any cutoff surface there are more $R^-$'s than $R$'s.  Were there equal number of them the above relations would hold; but $M$'s and $\mu$ give probabilities on different sample spaces so this naive expectation is wrong.

Let us see this in a simple example.  Suppose that there are only two possible outcomes: 1 and 2.  We first calculate regularized $M(1,1)$ under a cutoff surface:
\begin{equation}
M(1,1) = \frac{N_\Diamond(1,1)}{N_\Diamond},
\end{equation}
where $N_\Diamond$ is the total number of regions $R$ and $N_\Diamond(1,1)$ is the number of such regions with initial outcome 1 and final outcome 1.  Similarly,
\begin{equation}
M(1|1) = \frac{N_\Diamond(1,1)}{N_\Diamond(1)}, \qquad \mu(1) = \frac{N_\vee(1)}{N_\vee},
\end{equation}
where $N_\vee$ is the total number of past boundaries $R^-$ and ``$(1)$'' means that only regions with initial outcome 1 are counted.  Clearly under any cutoff surface $N_\Diamond \leq N_\vee$.  We find from these relations that:
\begin{equation}
M(1,1) = \frac{N_\vee / N_\Diamond}{N_\vee(1) / N_\Diamond(1)} \cdot \mu(1) M(1|1),
\end{equation}
which violates \eq{M=muM} unless in the limit that the cutoff goes to infinity the fraction is equal to 1, i.e., if
\begin{equation}\label{cond}
\frac{N_\vee(1)}{N_\Diamond(1)} = \frac{N_\vee(2)}{N_\Diamond(2)}.
\end{equation}
But this can fail to be true if, for some reason, the rate of progress of the cutoff surface is correlated with regions of high abundance of initial outcome 1 (or 2).

As a simple example, imagine a multiverse with only two types of vacua with different cosmological constants $H_{1,2}$.  We employ the scale factor time cutoff $t$ which is related to the proper time $\tau$ via $dt=H d\tau$.  Suppose that the double experiment we just mentioned consists of measuring the cosmological constant twice in a proper time interval $\ell$, with $n_{1,2}$ being the number density of experiments in their respective vacua.  Neglecting the remote possibility of a vacuum decay amid the experiment, we have a trivial conditional probability: $M(1|1)=M(2|2)=1$ and $M(1|2)=M(2|1)=0$.  Then if \eq{M=muM} held, we would have $M(1,1)=\mu(1)$ and $M(2,2)=\mu(2)$, or by dividing them:
\begin{equation}\label{condEx}
\frac{N_\vee(1)}{N_\vee(2)} = \frac{N_\Diamond(1,1)}{N_\Diamond(2,2)},
\end{equation}
which is equivalent to \eq{cond}.  We now show explicitly that this is violated in our example.  At late times the regularized number $N^t_\vee(i) = n_i V(t)$ where $V(t) = V_i(0) e^{\gamma t}$ is the asymptotic behavior of the volume ($\gamma$ and $V(0)$ are determined by the greatest eigenvalue and eigenvector of the matrix describing the exponential growth of volume and tunneling between vacua; see \cite{Linde94}).  Since the double experiment takes a proper time $\ell$ to complete, which is equal to a scale factor time $H_i\ell$, we have $N^t_\Diamond(i,i)=N^{t-H_i\ell}_\vee(i)$.  We therefore find that the RHS of \eq{condEx} has an extra factor $e^{\gamma\ell(H_2-H_1)}$ that's missing in the LHS.  Therefore \eq{cond}, and hence \eq{M=muM} are violated in our example.

Although most cutoff measures are vulnerable to this problem, a few exceptions are worth to mention.  First, the proper time cutoff measure is not affected simply because the {\it proper} time duration of the experiments are the same regardless of what vacuum they are located in.  Alternatively, it can be easily checked that \eq{cond} is satisfied if proper time is used.  Second, the stationary measure does not suffer from this problem because different cutoff surfaces are used in making predictions for events with different stationarity times and so \eq{cond} is not relevant.  Finally, the causal diamond measure is not affected by this problem, as it's stated here.  (The causal diamond measure is not a global time cutoff measure so our argument leading to \eq{cond} does not apply.)  This is because when $R^-$ falls below the regulating diamond then $R^+$ will fall there as well.  However, if we consider the example of a hyper-rectangle region $R$ with shielding walls we can break it into $R^+$ (the future spacelike surface), $R^-_i$ (the past spacelike surface) and $R^-_w$ (the six timelike walls).  Now we can ask the same questions about the abundance of $\phi_1$ and $\phi_2$ but on $R^-_i$ and $R^+$, respectively.  Then the causal diamond measure runs into trouble because the walls can cross the regulating diamond.

It should be emphasized that there is no mathematical contradiction here; $\mu$ and $M$'s are three independent probability distributions that are not related to each other through Eqs.~(\ref{M=muM}) and (\ref{mu=intM}).  However, if we use these three functions as predictions for the results of local experiments, then that would contradict basic properties of probability theory that we expect to hold for experimental probabilities.

\section{Local Experiments}

In this section we review local experiments and the process by which we can experimentally extract information from measurements.  Unless explicitly stated, our discussion does not depend on whether the system is classical or quantum.  A local observer performs experiments in a local lab and records the outcomes $\phi$ of the measurement.  We assume that the same measurement can be done independently on copies of a system to yield probabilities on the possible outcomes.  To begin, suppose that initially at $t_1=0$ the observer knows nothing about the state of the system, i.e., there is no previous measurement.  Then he conducts a measurement and records the experimental probability distribution $W(\phi; t_1=0)$.  He can perform a second measurement later and record the conditional probability $W(\phi_2; t_2 | \phi_1; t_1=0)$.  We will suppress the time label hereafter but we always take the time of the first measurement to be $t_1=0$.  More measurements can be done in a similar fashion.\footnote{It is important that the $R^+$ surface of each measurement constitutes the $R^-$ of the next one.  Equivalently, we can say that the lab is shielded from the external effects so that experiments are not affected by any causal signal besides those of the previous ones.}  Therefore the most general information an observer can acquire from successive measurements can be captured by weight functions of the form $W(\phi_n | \phi_{n-1}, \cdots, \phi_1)$, which satisfy the obvious normalization condition:
\begin{equation} \label{normW}
\int d\phi_n W(\phi_n | \phi_{n-1}, \cdots, \phi_1) = 1.
\end{equation}

It is possible to convey the same information via joint probability distributions.  For example and for future reference, we can define the joint $W(\cdot,\cdot)$ for two measurements by
\begin{equation} \label{W=WW}
W(\phi_2, \phi_1) = W(\phi_1) W(\phi_2 | \phi_1),
\end{equation}
which has a different normalization condition:
\begin{equation}
\int d\phi_1 d\phi_2 W(\phi_2, \phi_1) = 1.
\end{equation}
Conversely, starting from a given joint $W(\cdot,\cdot)$ we can define
\begin{equation} \label{W=intW}
W(\phi_1) = \int d\phi_2 W(\phi_2, \phi_1)
\end{equation}
and $W(\phi_2 | \phi_1) = W(\phi_2, \phi_1) / W(\phi_1)$ which satisfy normalization conditions of the form (\ref{normW}).  This establishes that the pair of functions $W(\cdot)$ and conditional $W(\cdot|\cdot)$ is equivalent to the single function joint $W(\cdot,\cdot)$.

Before proceeding, note that $\phi$ labels all possible outcomes of a measurement.  Therefore $W$ depends on the sequence ${\cal E}_n$ of measurements performed at each step, although this is not explicit in our notation.  So, for example, if we first measure the position and then the momentum of a particle we have $W(p_2 | x_1)$, whereas if the second measurement is also a position measurement then we have $W(x_2 | x_1)$.  Here $\phi$ is either $x$ or $p$.

To motivate the forthcoming definitions, consider an ensemble of experimenters each performing three measurements ${\cal E}_{1,2,3}$ at times $t_{1,2,3}$.  Now suppose that they find the empirical probability $W(\phi_3|\phi_2,\phi_1)$ to be independent of the outcome $\phi_1$ of the first measurement.  It then may occur to the experimenters that $\phi_2$ is sufficient for predicting $\phi_3$.  To confirm this they would vary the type of the first measurement ${\cal E}_1$ and see if the result is still independent of $\phi_1$.  They would further need to add additional prior measurements and check if $W(\phi_3|\phi_2,\phi_1,\cdots)$ is independent of all these additional outcomes.  Once convinced of sufficiency of $\phi_2$ for predicting $\phi_3$ (in other words, irrelevance of $\phi_1, \cdots$) they can claim that $\phi_2$ determines the state of the system.  We can generalize these statements when $\phi_2$ is a series of measurements that together determine the state.  To give an example, classical mechanics asserts that $\phi_2=(x,p)$ (or $(p,x)$---the order doesn't matter, but they have to be at the same time $t_2$) determines the state of a single particle and $W(\phi|(x,p),\cdots)$ is independent of everything prior to the measurement of $(x,p)$.  Now we can abstract the notion of Markov property from this example.

We say that $W$ has a ``Markov property'' (in time) associated to a sequence $({\cal E}_i)_{i=1}^n$ of immediately successive experiments, if all experimental probabilities $W(\phi | \phi_{m+n}, \cdots, \phi_{m+1}, \cdots)$ are independent of the number $m$, type  and outcome of measurements prior to $t_{m+1}$ given that $(t_{m+n} - t_{m+1})$ is sufficiently small.  Here $\phi_{m+i}$ is the outcome of a measurement of type ${\cal E}_i$ (but the rest of the measurements are arbitrary).  We call a minimal such $({\cal E}_i)_1^n$ a ``complete set of measurements'' and say that its outcome $\Phi = (\phi_i)_1^n$ gives the ``complete state'' (or ``state,'' for short) of the system after the last measurement ${\cal E}_n$.\footnote{Note that with this definition there may exist ``hidden variables'' that the experimentalist is blind to.  For example, if one studies an ideal gas and chooses to measure \emph{only} $P$, $V$ and $T$ then $(P,V)$ determines the state of the system.  He is simply blind to atomic velocities and positions that, according to statistical mechanics, constitute the state of the system.  However, a thermodynamicist would agree with him on calling $(P,V)$ the state of the system.}  We refer to any subset of a complete state as an incomplete state (but employ the same symbol $\phi$ as for a single outcome).  We further assume that within a complete set of measurements the outcomes of incomplete measurements are independent of each other (commuting, in the context of quantum mechanics).

With this notation we can write:
\begin{equation} \label{Markov}
W(\phi_n | \Phi_{n-1}, \cdots) = W(\phi_n | \Phi_{n-1}).
\end{equation}
It follows that as far as the future evolution is concerned all that matters is the initial state of the system, known at the last complete measurement, not how it got to that state:
\begin{equation} \label{genMarkov}
W(\phi_n | \phi_{n-1}, \cdots, \Phi_i, \cdots) = W(\phi_n | \phi_{n-1}, \cdots, \Phi_i).
\end{equation}
Note that if $\Phi_1 \neq \Phi_2$ are obtained from the same set of complete measurements, then they describe distinct states.  But it is possible that two sets of complete measurements lead to two totally different, but equivalent, sets of outcomes $\Phi$ and $\Phi'$ which describe the same state.  $\Phi$ and $\Phi'$ are equivalent if they lead to identical subsequent evolutions:
\begin{equation}
W(\phi | \Phi) = W(\phi | \Phi'), \quad \hbox{for all $\phi$.}
\end{equation}
The set of equivalence classes of $\Phi$'s defines the space of states of the system.

The preceding definition of state was from the point of view of an experimentalist who has access to $W$.  From the theoretical perspective, states and their Markovian evolution play a central role and $W$ is derived from them as a prediction.  This leads to numerous correlations among $W$'s and hence to a huge reduction in the information they carry.  Let us review how it works.  In classical mechanics the state of the system is a point $(x,p)$ in its phase space which evolves according to a Hamiltonian.  Everything is deterministic except the initial state (just before the first measurement) which is determined by a probability distribution $P(x,p)$.  The result of all other experiments are found from the probability distribution induced on the measured observable.  In quantum mechanics of a closed system the state of the system is a vector in its Hilbert space.  Again the evolution is governed by a Hamiltonian, but the outcomes of measurements are not deterministic: $\lim_{t_2 \to t_1} W(\Phi_2 | \Phi_1) = |\langle \Phi_2 | \Phi_1 \rangle|^2$, where $| \Phi_{1,2} \rangle$ are vectors corresponding to the states $\Phi_{1,2}$.  The initial state (just before the first measurement) is determined by a probability distribution on the Hilbert space which is practically equivalent to a density matrix: $W(\Phi) = \langle \Phi | \rho | \Phi \rangle$.

In the next section we use these properties of $W$, especially the Markov property, to address the problem encountered in the previous section.

\section{Geocentric Measures}

We can now summarize the problem with cutoff measures we found above:  If we identify
\bea
W(\phi) \equiv \mu(\phi), \\
\quad W(\phi_2 | \phi_1) \equiv M(\phi_2 | \phi_1), \\
\quad W(\phi_2, \phi_1) \equiv M(\phi_2, \phi_1),
\eea
then Eqs.~(\ref{W=WW}) and (\ref{W=intW}) imply Eqs.~(\ref{M=muM}) and (\ref{mu=intM}), but the latter are not necessarily satisfied for all cutoff measures.  One way to escape this puzzle is to count only those initial outcomes for which the whole experiment is under the cutoff: $\mu(1) = N_\Diamond(1)/N_\Diamond$.  This is equivalent to discarding the $\mu$ computed by the measure completely and deriving everything from the joint $M(\cdot,\cdot)$.  In particular $\mu$ is now defined by \eq{mu=intM} which automatically fixes the problem.  However, this would lead to predictions for local experiments which differ from the known local physics. In addition, for variable-time experiments $M(\cdot,\cdot)$ can still suffer from paradoxes \cite{GV}. So we will not pursue this solution here. 

A different perspective on the problem of cutoff measures we discussed is that this puzzle stems from the insufficiencies of semi-classical description of quantum cosmology and that we are using the measure more than once (see Appendix B in \cite{LVW}).  This approach is closely related to what we are going to propose in this Section where the cutoff measure is invoked, if at all, only once. Contrary to the solution of the preceding paragraph, we can simply forget about computing $M(\cdot,\cdot)$ and $M(\cdot|\cdot)$ from the measure and be content with what we get for $\mu$. We can define $M(\cdot,\cdot)$ such that \eq{M=muM} holds but we still need $M(\cdot|\cdot)$.  This motivates our approach in this section where we derive all probabilities from an \emph{initial state probability} (corresponding to $\mu$) and some \emph{evolution or transition probability} (corresponding to conditional $M(\cdot|\cdot)$).  We study these generically and will not confine ourselves to $\mu$'s that are derived from the known cutoff measures.

The Markov property (\ref{Markov}) suggests that a whole set of $W$'s can be produced by only two functions:
\be
\mu(\Phi) = W(\Phi), \qquad
\ee
 and
 \be
 \qquad P(\Phi_2 | \Phi_1) = W(\Phi_2; t_2=T | \Phi_1; t_1=0),
\ee
which we call the ``measure'' $\mu$ and the ``local transition probability'' $P$.

In most physical theories, the local transition probability is determined by a unitary evolution via a Hamiltonian.  But in this paper we are not interested in properties of $P$ (which we assume to be given) and would like to study $\mu$.  It is clear that the shielded experiments that we discussed earlier do not help.  Instead we have to let in external signals from outside of the already-explored initial surface $R^-$ to probe information about $\mu$ that has not yet been accessible.

So far we have only made use of the Markov property of $W$ in time to break it into $\mu$ and $P$.  As mentioned before, according to quantum mechanics the most general $\mu$ is a density matrix.  We restrict to a smaller class of $\mu$'s by considering a classical field whose state is given by $\varphi(x)$ and its conjugate momentum $\pi(x)$ defined on the initial surface $R^-$.  These field configurations correspond to coherent states in the quantum theory and hence reduce the density matrix to a functional $\mu[\varphi,\pi]$.  Now we impose an additional constraint on $\mu$ by demanding that the probabilities for field values in a region depend on those outside only through their common boundary.  We shall refer to this assumption as the Markov property in space.  This assumption allows us to find the probability distribution everywhere by propagating our knowledge outward, hence possible names lab-centric, observer-centric or simply {\it geocentric}.

Let us briefly recall some basic facts of the theory of Markov random fields.  Assume a set of random variables that are interconnected by a graph.  A Markov random field is this set of variables with the property that given the neighbors of any of its variables, it is conditionally independent from all other variables.  Roughly speaking, the edges of the graph determine which random variables are directly dependent.  The graph can be thought of as the union of (potentially overlapping) cliques.  (A clique is a complete subgraph where any pair of vertices are connected by an edge).  The Hammersley-Clifford theorem \cite{HC} states that the joint probability density of the random variables factorizes over the cliques, i.e., it can be written as a product of functions each dependent only on vertices of a single clique.

Using this terminology we can sharpen our definition of Markov property in space.  In our case, the Markov random field is the set $\{\varphi(x),\pi(x) | \forall x\}$ in a discretized space.  The graph connects every point (i.e. $\varphi$ and $\pi$ at every point) to its nearest neighbors on a cubic lattice as well as to its diagonal neighbors.  Then the Hammersley-Clifford theorem implies that we can write $\mu$ as the exponentiation of a local functional of fields:\footnote{The form of the `Lagrangian' in the exponent depends on the graph.  If, for example, diagonal neighbors were removed then we wouldn't have product of field derivatives in orthogonal directions.}
\begin{equation}
\mu[\varphi, \pi] = \exp \left[ -\int_{R^-} d^3x {\cal L} \left( \varphi(x), \nabla \varphi(x); \pi(x), \nabla \pi(x) \right) \right].
\label{eq:path_integral}
\end{equation}
We have used the symbol $\cal L$ to draw an analogy between $\mu$ and $\cal L$ on one hand and the partition function and Lagrangian density on the other hand.  Therefore, we see that with the assumptions of Markov property in space and time the space of measures on classical fields is equivalent to the space of 3D Lagrangians, and the search for a measure is completely equivalent to a search for a 3D Lagrangian. 

As an example, consider a classical statistical mechanics of electromagnetic field at finite temperature $T$. In temporal gauge the partition function is given by
\be
Z = \int e^{- \frac{1}{T} \int d^3 x {\cal H}(P, A)} {\cal D} {A} {\cal D} {P}, 
\label{eq:partition}
\ee
where $P_i \equiv \dot{A}_i$, $F_{ij} \equiv \partial_i A_j - \partial_j A_i$ and ${\cal H}= \frac{1}{4} F^2_{i j} + \frac{1}{2}P^2_i$.
It follows that the corresponding 3D Lagrangian for a thermal bath of photons could be written as
\be
{\cal L} = \frac{\cal H}{T} = \frac{1}{4 T} F^2_{i j} + \frac{1}{2 T}P^2_i.
\label{eq:Lagrangian}
\ee
Of course additional constraints must be added to the 3D path integrals of Eqs.~(\ref{eq:path_integral}) and (\ref{eq:partition}) to impose the Gauss's law. Therefore a geocentric measure determined by Eq.~(\ref{eq:Lagrangian}) can describe a thermal CMB radiation with $T \approx 2.7 K$ given that appropriate additional fields and/or interactions are added to explain, for example, the acoustic peaks.

It is also possible to obtain a consistent measure from any of the cutoff measures by using them only once to compute $\mu$ on the initial surface at the time of the first measurement. As was argued above for the scale factor cutoff measure \eq{cond} and hence \eq{M=muM} are not satisfied if ``1'' and ``2'' refer to high and low cosmological constant.  But if we use this measure only to compute $\mu$ but not $M(\cdot|\cdot)$ then it is consistent but may or may not have the Markov property in space.

\section{Conclusion} 

We have shown that most of the global time cutoff measures of the multiverse suffer from severe inconsistencies and developed a new framework which allows us to study the measure problem from a completely different perspective.  In the emerging picture an infinite multiverse is replaced with a finite geocentric region, and the search for the correct measure is replaced by a search for a 3D Lagrangian yet to be discovered. 

There are two ways to look for the correct Lagrangian. One could either try to perform direct phenomenological searches or one could try to derive it from first principles. For the phenomenological approach one has to reinterpret the existing cosmological data from the geocentric view point. Although we are guaranteed to uncover some Lagrangian it is not a priori guaranteed that the corresponding Lagrangian will be local, simple or even useful. On the other hand it would be interesting to see whether one can derive the corresponding theory from some yet-unknown first principles.

\section*{Acknowledgments}

The authors are grateful to Andrei Linde for encouragement and very useful conversations, and to Alexey Golovnev, Alan Guth, Daniel Harlow, Dusan Simic, Lenny Susskind and Alex Vilenkin for helpful discussions. This work was supported in part by NSF Grant No.~0756174. V.V.\ was also supported by FQXi mini-grants MGB-07-018 and MGA-09-017. M.N. was supported by Mellam Family Foundation.

\end{document}